\documentclass[sigconf]{acmart}

\usepackage{todonotes}
\usepackage{enumitem}
\usepackage{soul}
\usepackage[linewidth=0.2pt]{mdframed}
\newcommand{\sm}[1]{#1}

\newcommand{\smrm}[1]{#1}
\newcommand{\smrmf}[1]{#1}
\newcommand{\smadd}[1]{}

\AtBeginDocument{%
  \providecommand\BibTeX{{%
    \normalfont B\kern-0.5em{\scshape i\kern-0.25em b}\kern-0.8em\TeX}}}

\copyrightyear{2020} 
\acmYear{2020} 
\setcopyright{acmcopyright}\acmConference[SIGIR '20]{Proceedings of the 43rd International ACM SIGIR Conference on Research and Development in Information Retrieval}{July 25--30, 2020}{Virtual Event, China}
\acmBooktitle{Proceedings of the 43rd International ACM SIGIR Conference on Research and Development in Information Retrieval (SIGIR '20), July 25--30, 2020, Virtual Event, China}
\acmPrice{15.00}
\acmDOI{10.1145/3397271.3401262}
\acmISBN{978-1-4503-8016-4/20/07}

\begin{document}

% \title[A Representation-Based Model of Term Salience for Passage Re-Ranking]{A Representation-Based Model of \\ Term Salience for Passage Re-Ranking}
\title[Expansion via Prediction of Importance with Contextualization]{Expansion via Prediction of Importance with Contextualization}
% \subtitle{A Contextualized Representation Approach for Passage Re-Ranking}

\author{Sean MacAvaney}
\affiliation{\institution{IR Lab, Georgetown University, USA}}
\email{sean@ir.cs.georgetown.edu}

\author{Franco Maria Nardini}
\affiliation{\institution{ISTI-CNR, Pisa, Italy}}
\email{francomaria.nardini@isti.cnr.it}

\author{Raffaele Perego}
\affiliation{\institution{ISTI-CNR, Pisa, Italy}}
\email{raffaele.perego@isti.cnr.it}

\author{Nicola Tonellotto}
\affiliation{\institution{University of Pisa, Italy}}
\email{nicola.tonellotto@unipi.it}

\author{Nazli Goharian}
\affiliation{\institution{IR Lab, Georgetown University, USA}}
\email{nazli@ir.cs.georgetown.edu}

\author{Ophir Frieder}
\affiliation{\institution{IR Lab, Georgetown University, USA}}
\email{ophir@ir.cs.georgetown.edu}

\renewcommand{\shortauthors}{MacAvaney et al.}

\fancyhead{}

\begin{abstract}
The identification of relevance with little textual context is a primary challenge in passage retrieval. We address this problem with a representation-based ranking approach that: (1) explicitly models the importance of each term using a contextualized language model; (2) performs passage expansion by propagating the importance to similar terms; and (3) grounds the representations in the lexicon, making them interpretable. Passage representations can be pre-computed at index time to reduce query-time latency. We call our approach EPIC (Expansion via Prediction of Importance with Contextualization). We show that EPIC significantly outperforms prior importance-modeling and document expansion approaches. We also observe that the performance is additive with the current leading first-stage retrieval methods, further narrowing the gap between inexpensive and cost-prohibitive passage ranking approaches. Specifically, EPIC achieves a MRR@10 of 0.304 on the MS-MARCO passage ranking dataset with 78ms average query latency on commodity hardware. We also find that the latency is further reduced to 68ms by pruning document representations, with virtually no difference in effectiveness.
\end{abstract}

\maketitle

%!TEX root = main.tex
\section{Introduction}

\begin{figure}
\centering
\vspace{1em}
\includegraphics[scale=0.83]{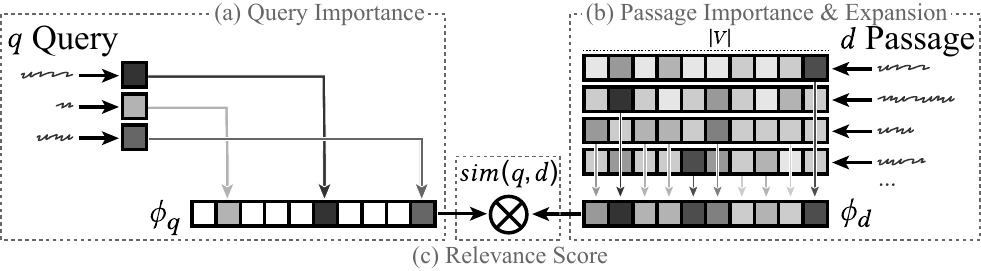}
\vspace{-0.5em}
\caption{Overview of EPIC.}
\vspace{-0.5em}
\label{fig:overview}
\end{figure}

Passage retrieval is fundamentally burdened by short passages. While document retrieval systems can rely on signals such as term frequency to estimate the importance of a given term in a document, passages usually do not have this benefit. Consequently, traditional retrieval approaches  often perform poorly at passage retrieval. Supervised deep learning approaches---in particular, those that make use of pretrained contextualized language models---have successfully overcome this limitation by making use of general language characteristics~\cite{Craswell2019OverviewTRECDL,Hashemi2019ANTIQUEAN}. However, these approaches have a substantial computational burden, which can make them impractical~\cite{Hofsttter2019LetsMR,nogueiradoc2query}.

We propose a new approach for passage retrieval that performs modeling of term importance (i.e., salience) and expansion over a contextualized language model to build query and document representations. We call this approach EPIC (Expansion via Prediction of Importance with Contextualization). At query time, EPIC can be employed as an inexpensive re-ranking method because document representations can be pre-computed at index time. EPIC improves upon the prior state of the art on the MS-MARCO passage ranking dataset by substantially narrowing the effectiveness gap between practical approaches with subsecond retrieval times and those that are considerably more expensive, e.g., those using BERT as a re-ranker.
Furthermore, the proposed representations are  interpretable because the dimensions of the representation directly correspond to the terms in the lexicon. An overview is shown in Fig.~\ref{fig:overview}.

Neural re-ranking approaches can generally be characterized as either representation-based or interaction-based~\cite{Guo2016ADR}. Representation-based models, like ours, build representations of a query and passage independently and then compare these representations to calculate a relevance score. These are beneficial because one can compute document representations at index time to reduce the query-time cost.
Interaction-based models combine signals from the query and the document at query time to compute the relevance score~\cite{nogueira2019passage}. The Duet model~\cite{mitra2019updated} aims to achieve low query-time latency by combining signals from both a representation-based and an interaction-based model.
However, this approach substantially under-performs the latest pure interaction-based approaches such as the one in~\cite{nogueira2019passage}. TK~\cite{hofstatter2020interpretable} attempts to bridge this performance gap by  using a smaller transformer network, but still utilizes an interaction-based approach which itself adds considerable computational overhead. Finally, other interesting proposals have investigated alternative approaches for offloading computational cost to index time. Doc2query~\cite{nogueira2019document} and docTTTTTquery~\cite{nogueiradoc2query} add important context to otherwise short documents by using a sequence-to-sequence model to predict additional terms to add to the document. DeepCT-Index~\cite{dai2019context} models an importance score for each term in the document and replaces the term frequency values in the inverted index with these scores.
Unlike these approaches, EPIC models query/document term importance and performs document expansion. We found that it can be employed as an inexpensive yet effective re-ranking model;
the impact on query time latency can be as low as an additional 5ms per query (for a total of 68ms).

In summary, the novel contributions presented are:
\begin{itemize}[leftmargin=*,topsep=2pt]
\item[-] We propose a new representation-based ranking model that is grounded in the lexicon.
\item[-] We show that this model can improve ranking effectiveness for passage ranking, with a minimal impact on query-time latency.
\item[-] We show that the model yields interpretable representations of both the query and the document.
\item[-] We show that latency and storage requirements of our approach can be reduced by pruning the document representations.
\item[-] For reproducibility, our code is integrated into OpenNIR~\cite{macavaney:wsdm2020-onir}, with instructions and trained models available at: \\ \url{https://github.com/Georgetown-IR-Lab/epic-neural-ir}.
\end{itemize}

%!TEX root = main.tex
\section{Model}

\textbf{Overview and notation.}
Our model follows the representation-focused neural ranking paradigm. That is, we train a model to generate a query and document\smadd{~(or passage)}\smrm{\footnote{For ease of notation, we refer to passages as documents.}} representation in a given fixed-length vector space, and produce a ranking score by computing a similarity score between the two representations.

Assume that queries and documents are composed by sequences of terms taken from a vocabulary $V$.
Any sequence of terms, either a query or a document, is firstly represented as a sequence of vectors using a contextualized language model like BERT~\cite{devlin-19}. More formally, let $f: V^n \to \mathbb{R}^{n \times e}$ denote such a function associating  an input sequence $s$ of $n$ terms $t_1, \ldots, t_n$ to their $n$ embeddings $f_1(s), \ldots, f_n(s)$, where $f_i(s) \in \mathbb{R}^e$ and $e$ is the size of the embedding. So, a $n$-term query $q$ is represented with the $n$ embeddings $f_1(q), \ldots, f_n(q)$, and a $m$-term document $d$ is represented with $m$ embeddings $f_1(d), \ldots, f_m(d)$. Given the embeddings for queries and documents, we now illustrate the process for constructing query representations, document representations, the final query-document similarity score.

\textbf{Query representation.} A query $q$ is represented as a \textit{sparse} vector $\phi_q \in \mathbb{R}^{|V|}$ \sm{(Fig.~\ref{fig:overview} (a))}. The elements of $\phi_q$ that correspond to to terms not in the query are set to $0$. For each term $t_i$ appearing in the $t_1, \ldots, t_n$ terms of the query $q$, the corresponding element $\phi_q(t_i)$ is equal to the importance $w_q(t_i)$ of the term w.r.t. the query
\begin{equation}\label{eq:querysaliency}
w_q(t_i) = \ln\Big(1+\mathrm{softplus}\big(\theta_1^\top f_i(q)\big)\Big),
\end{equation}
where $\theta_1 \in \mathbb{R}^e$ is a vector of learned parameters. The $\mathrm{softplus}(\cdot)$ function is defined as $\mathrm{softplus}(x)=\ln(1+e^x)$. The use of softplus ensures that no terms have a negative importance score, while imposing no upper bound. The logarithm prevents individual terms from dominating.
When a term appears more than once in a query, the corresponding value of $\phi_q$ sums up all contributions. The elements of the query representation encode the importance of the terms w.r.t. the query. This approach allows the query representation model to learn to assign higher weights to the query terms that are most important to match given the textual context. Note that the number of elements in the representation is equal to the number of query terms; thus the query processing time is proportional to the number of query terms~\cite{fntir2018}.

\textbf{Document representation.} A document $d$ is represented as a \textit{dense} vector $\phi_d \in \mathbb{R}^{|V|}$ \sm{(Fig.~\ref{fig:overview} (b))}.
Firstly, to perform document expansion, each $e$-dimensional term embedding $f_j(d)$ is projected into a $|V|$-dimensional vector space, i.e., $\psi_j: f_j(d) \mapsto \Theta_2 f_j(d)$, where $\Theta_2 \in \mathbb{R}^{|V| \times e}$ is a matrix of learned parameters. Note that $\psi_j \in \mathbb{R}^{|V|}$, and let $\psi_j(\tau)$ denote the entry of this vector corresponding to term $\tau \in V$. Secondly, the importance $w_d(t_j)$ of the terms w.r.t. the document is computed as in Eq~\eqref{eq:querysaliency}:
\begin{equation}\label{eq:docsaliency}
	w_d(t_j) = \ln\Big(1+\mathrm{softplus}\big(\theta_3^\top f_j(d)\big)\Big),
\end{equation}
where $\theta_3 \in \mathbb{R}^e$ is a vector of learned parameters. Thirdly, we compute a factor representing the overall quality $c(d)$ of the document
\begin{equation}
	c(d) = \mathrm{sigmoid}(\theta_4^\top d_{\texttt{[CLS]}}),
\end{equation}
where $\theta_4 \in \mathbb{R}^e$ is a vector of learned parameters, and $d_{\texttt{[CLS]}} \in \mathbb{R}^e$ is the embedding produced by the contextualized language model's classification mechanism. We find that this factor helps give poor-quality documents lower values overall. The $\mathrm{sigmoid}(\cdot)$ function is defined as: $\mathrm{sigmoid}(x)=\frac{1}{1+e^{-x}}$.
Finally, for each term $\tau$ appearing in the vocabulary, the corresponding element of the document representation $\phi_d(\tau)$ is defined~as:
\begin{equation}
	\phi_d(\tau) = c(d) \max_{t_j \in d}\big(w_d(t_j)\psi_j(\tau)\big).
\end{equation}
This step takes the maximum score for each term in the vocabulary generated by any term in the document. Since they do not rely on the query, these representations can be computed at index time.

\textbf{Similarity measure.} We use the dot product to compute the similarity between the query and document vectors \sm{(Fig.~\ref{fig:overview} (c))}, i.e.,
\begin{equation}
sim(q,d)= \phi_q^\top \phi_d = \sum_{\tau \in V} \phi_q(\tau) \phi_d(\tau).
\end{equation}

\section{Experimental Evaluation}
\label{sec:experiments}

We conduct experiments using the MS-MARCO passage ranking dataset (full ranking setting).\smrm{\footnote{\url{https://microsoft.github.io/msmarco/}}} This dataset consists of approximately $1$ million natural-language questions gathered from a query log (average length: 7.5 terms, stddev: 3.1), and $8.8$ million candidate answer passages (avg length: 73.1, stddev: 28.4). The dataset is shallowly-annotated. Annotators were asked to write a natural-language answer to the given question using a set of candidate answers from a commercial search engine. The annotators were asked to indicate which (if any) of the passages contributed to their answers, which are then treated as relevant to the question. This results in 0.7 judgments per query on average (1.1 judgments per query of the 62\% that have an answer). Thus, this dataset has a lot of variation in queries, making it suitable for training neural ranking methods. Although this dataset is limited by the method of construction, the performance on these shallow judgments correlate well with those conducted on a deeply-judged subset~\cite{Craswell2019OverviewTRECDL}.

\textbf{Training}. We train our model using the official MS-MARCO sequence of training triples (query, relevant passage, presumed non-relevant passage) using cross-entropy loss. We use BERT-base~\cite{devlin-19} as the contextualized language model, as it was shown to be an effective foundation for various ranking techniques~\cite{dai2019context,macavaney:sigir2019-cedr,nogueira2019passage}. We set the dimensionality $|V|$ of our representations to the size of the BERT-base word-piece vocabulary ($d$=30,522). The embedding size is instead $e=768$. $\Theta_2$ is initialized to the pre-trained masked language model prediction matrix; all other added parameters are randomly initialized. Errors are back-propagated through the entire BERT model with a learning rate of $2\times10^{-5}$ with the Adam optimizer~\cite{Kingma2015AdamAM}. We train in batches of 16 triples using gradient accumulation, and we evaluate the model on a validation set of 200 random queries from the development set every 512 triples. The optimal training iteration and re-ranking cutoff threshold is selected using this validation set. We roll back to the top-performing model after 20 consecutive iterations (training iteration 42) without improvement to Mean Reciprocal Rank at 10 (MRR@10).

\begin{table}
\centering
\caption{Effectiveness and efficiency of our approach compared to a variety of baselines. The values in \textit{italics} represent a good trade-off between effectiveness and query latency. The value marked with~* was reported by~\cite{nogueiradoc2query}.}
\label{tab:results}
{
\renewcommand{\arraystretch}{0.9}
\begin{tabular}{lrr}
\toprule
& MS-Marco Dev & Latency \\
Method & MRR@10 & ms/query \\
\midrule
\textbf{Single-Stage Ranking}\\
BM25 (from Anserini~\cite{Yang2018AnseriniRR}) & 0.198 & 21\phantom{*} \\
doc2query~\cite{nogueira2019document} & 0.218 & 48\phantom{*} \\
DeepCT-Index~\cite{dai2019context} & 0.243 & 15\phantom{*} \\
docTTTTTquery~\cite{nogueiradoc2query} & 0.277 & 63\phantom{*} \\
\midrule
\multicolumn{3}{l}{\textbf{Representation-based Re-Ranking}}\\
EPIC + BM25 (ours) & 0.273 & 106\phantom{*} \\
\phantom{ - }pruned $r=2000$ & 0.273 & 104\phantom{*} \\
\it\phantom{ - }pruned $r=1000$ & \it 0.272 & \it 48\phantom{*} \\
EPIC + docTTTTTquery (ours) & 0.304 & 78\phantom{*} \\
\phantom{ - }pruned $r=2000$ & 0.304 & 77\phantom{*} \\
\it\phantom{ - }pruned $r=1000$ & \it 0.303 & \it 68\phantom{*} \\
\midrule
\multicolumn{3}{l}{\textbf{Other Re-Ranking}}\\
Duet (v2, ensemble)~\cite{mitra2019updated} & 0.252 & 440\phantom{*} \\
BM25 + TK (1 layer)~\cite{hofstatter2020interpretable} & 0.303 & 445\phantom{*} \\
BM25 + TK (3 layers)~\cite{hofstatter2020interpretable} & 0.314 & 640\phantom{*} \\
BM25 + BERT (large)~\cite{nogueira2019passage} & 0.365 & 3,500* \\
\bottomrule
\end{tabular}
}
\end{table}

\textbf{Baselines and Evaluation}. We test our approach by re-ranking the results from several first-stage rankers. We report the performance using MRR@10, the official evaluation metric, on the MS-MARCO passage ranking Dev set. We measure significance using a paired t-test at $p<0.01$. We compare the performance of our approach with the following baselines:
\begin{itemize}[leftmargin=*,topsep=2pt]
\item[-] BM25 retrieval from a Porter-stemmed Anserini~\cite{Yang2018AnseriniRR} index using default settings.\smrm{\footnote{We observe that the default settings outperform the BM25 results reported elsewhere and on the official leaderboard (e.g.,~\cite{nogueira2019document}).}}
\item[-] DeepCT-Index~\cite{dai2019context}, a model which predicts document term importance scores, and replaces the term frequency values with these importance scores for first-stage retrieval.
\item[-] doc2query~\cite{nogueira2019document} \sm{and docTTTTTquery~\cite{nogueiradoc2query}}, document expansion approaches which predict additional terms to add to the document via a sequence-to-sequence transformer model. These terms are then indexed and used for retrieval using BM25. \sm{The docTTTTTquery model uses a pre-trained T5 model~\cite{Raffel2019ExploringTL}.}
\item[-] Duet~\cite{mitra2019updated}, a hybrid representation- and interaction-focused model. We include the top Duet variant on the MS-MARCO leaderboard (version 2, ensemble) to compare with another model that utilizes query and document representations.
\item[-] TK~\cite{hofstatter2020interpretable}, a contextualized interaction-based model, focused on minimizing query time. We report results from~\cite{hofstatter2020interpretable} with the optimal re-ranking threshold and measure end-to-end latency on our hardware.
\item[-] BERT Large~\cite{nogueira2019passage}, an expensive contextualized language model-based re-ranker. This approach differs from ours in that it models the query and passage jointly at query time, and uses the model's classification mechanism for ranking.
\end{itemize}
We also measure query latency over the entire retrieval and re-ranking process. The experiments were conducted on commodity hardware equipped with an AMD Ryzen 3.9GHz processor, 64GiB DDR4 memory, a GeForce GTX 1080ti GPU, and a SSD drive.
We report the latency of each method as the average execution time (in milliseconds) of 1000 queries from the Dev set after an initial 1000 queries is used to warm up the cache. First-stage retrieval is conducted with Anserini~\cite{Yang2018AnseriniRR}.

\textbf{Ranking effectiveness}. We report the effectiveness of our approach in terms of MMR@10 in Table~\ref{tab:results}. When re-ranking BM25 results, our approach substantially outperforms doc2query and DeepCT-Index. Moreover, it performs comparably to docTTTTTquery (0.273 compared to 0.277, no statistically significant difference). More importantly, we observe that the improvements of our approach and docTTTTTquery are additive as we achieve a MRR@10 of 0.304 when used in combination. This is a statistically significant improvement, and substantially narrows the gap between approaches with low query-time latency and those that trade off latency of effectiveness (e.g., BERT Large).

To test whether EPIC is effective on other passage ranking tasks as well, we test on the TREC CAR passage ranking benchmark~\cite{Dietz2017TRECCA}. When trained and evaluated on the 2017 dataset (automatic judgments) with BM25, the MRR increases from 0.235 to 0.353. This also outperforms the DeepCT performance reported by~\cite{dai2019context} of 0.332.

\begin{figure}
\centering
\includegraphics[scale=0.55]{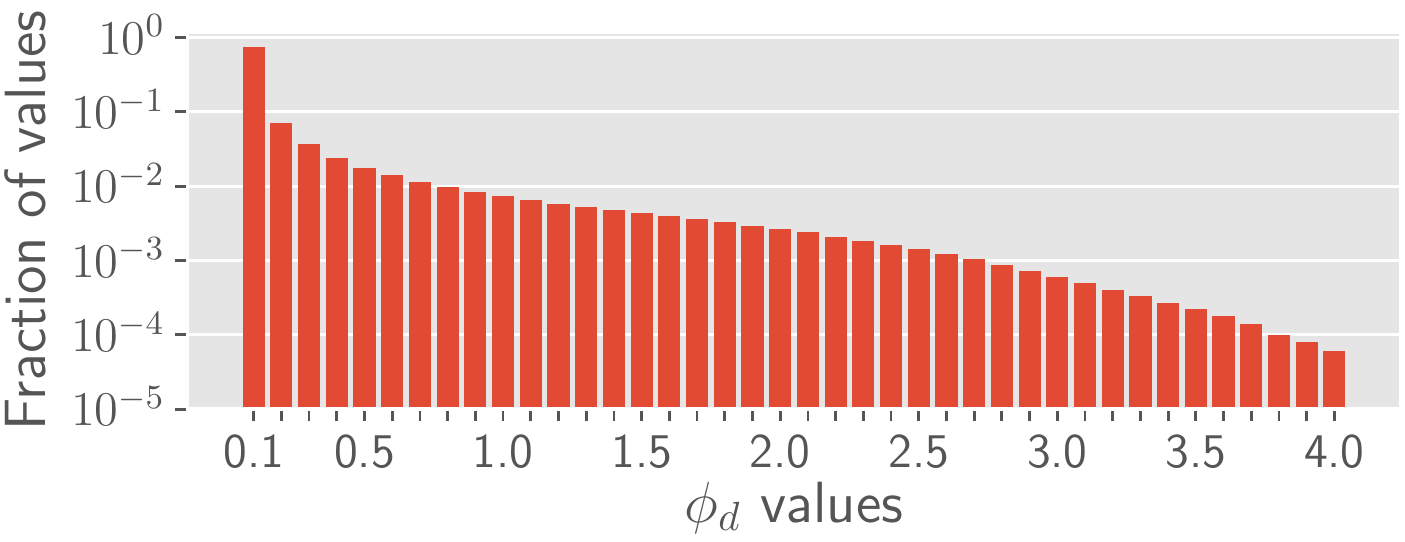}
\vspace{-1em}
\caption{Frequencies of values (log scale) appearing in the document representations. In this figure, values are rounded up to the nearest decimal value.}
\label{fig:dvec_freq}
\end{figure}

\textbf{Effect of document representation pruning.} For document vectors, we observe that the vast majority of values are very low (approximately 74\% have a value of 0.1 or below, see Fig.~\ref{fig:dvec_freq}). This suggests that many of the values can be pruned with little impact on the overall performance. This is desirable because pruning can substantially reduce the storage required for the document representations. To test this, we apply our method keeping only the top $r$ values for each document. We show the effectiveness and efficiency of $r=2000$ (reduces vocabulary by 93.4\%) and $r=1000$ (96.7\%) in Table~\ref{tab:results}. We observe that the vectors can be pruned to $r=1000$ with virtually no difference in ranking effectiveness (differences not statistically significant). We also tested with lower values of $r$, but found that the effectiveness drops off considerably by $r=100$ (0.241 and 0.285 for BM25 and docTTTTTquery, respectively).

\textbf{Ranking efficiency}. We find that EPIC can be implemented with a minimal impact on query-time latency. On average, the computation of the query representation takes 18ms on GPU and 51ms on CPU. Since this initial stage retrieval does not use our query representation, it is computed in parallel with the initial retrieval, which reduces the impact on latency. The similarity measure consistently takes approximately 1ms per query (both on CPU and GPU), with the remainder of the time spent retrieving document representations from disk. Interestingly, we observe that the latency of EPIC BM25 is higher than EPIC docTTTTTquery. This is because when re-ranking docTTTTTquery results, a lower re-ranking cutoff threshold is needed than for BM25. \smrm{This further underscores the importance of using an effective first-stage ranker.} When using pruning at $r=1000$, the computational overhead can be substantially reduced. Specifically, we find that EPIC only adds a 5ms overhead per query to docTTTTTquery, while yielding a significant improvement in effectiveness. With pruning at $r=1000$, EPIC BM25 performs comparably with docTTTTTquery with a $1.3\times$ speedup.

\textbf{Cost of pre-computing}. We find that document vectors can be pre-computed for the MS-MARCO collection in approximately $14$ hours on a single commodity GPU (GeForce GTX 1080ti). This is considerably less expensive than docTTTTTquery, which takes approximately $40$ hours on a Google TPU (v3). \smrm{When stored as half-precision (16-bit) floating point values, the vector for each document uses approximately 60KiB, regardless of the document length. This results in a total storage burden of approximately 500GiB for the entire collection. Pruning the collection to $r=1000$ (which has minimal impact on ranking effectiveness) reduces the storage burden of each document to 3.9KiB (using 16-bit integer indices) and total storage to 34 GiB.} \smadd{The pruned document vectors use 34 GiB, when stored as 16-bit floating point values.}

\begin{figure}
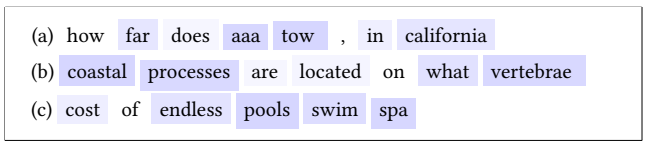

\small
\begin{flushleft}
\begin{mdframed}

(a)
\colorbox{blue!0}{how}	\colorbox{blue!9.85365853658538}{far}	\colorbox{blue!3.44715447154478}{does}	\colorbox{blue!15.7723577235773}{aaa}	\colorbox{blue!16}{tow }	\colorbox{blue!0.130081300812994}{,}    \colorbox{blue!6.30894308943093}{in}	\colorbox{blue!10.1788617886179}{california}

(b)
\colorbox{blue!16}{coastal}	\colorbox{blue!14.8674388674389}{processes}	\colorbox{blue!3.52123552123554}{are}	\colorbox{blue!3.89189189189196}{located}	\colorbox{blue!0}{on}	\colorbox{blue!11.8610038610039}{what}	\colorbox{blue!15.1763191763192}{vertebrae }

\smrm{
(c)
\colorbox{blue!6.64235624123419}{cost}	\colorbox{blue!0}{of}	\colorbox{blue!10.7938288920056}{endless}	\colorbox{blue!16}{pools}	\colorbox{blue!13.3295932678822}{swim}	\colorbox{blue!15.6858345021038}{spa}
}

\end{mdframed}
\end{flushleft}
\vspace{-1em}
\caption{Relative importance scores of sample queries. Darker colors correspond to higher weights\smrm{in the query representation}.}
\vspace{-0.5em}
\label{fig:qvec_interpret}
\end{figure}

\begin{figure}
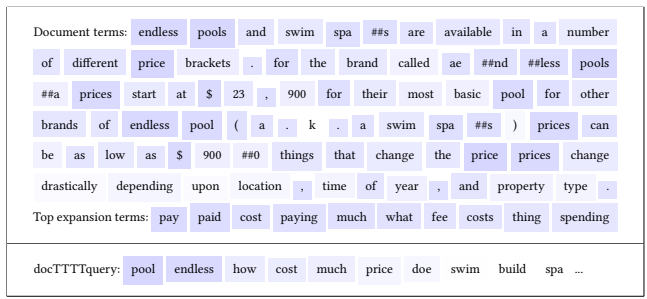

\tiny
\begin{flushleft}
\begin{mdframed}

Document terms:
\colorbox{blue!14.7510229776519}{endless}
\colorbox{blue!15.1740635819956}{pools}
\colorbox{blue!10.1076487252125}{and}
\colorbox{blue!8.8385269121813}{swim}
\colorbox{blue!12.3537928863708}{spa}
\colorbox{blue!12.5451683978596}{\#\#s}
\colorbox{blue!10.5206169342147}{are}
\colorbox{blue!9.45294302801385}{available}
\colorbox{blue!10.6213408876298}{in}
\colorbox{blue!10.8530059804847}{a}
\colorbox{blue!7.91186654076173}{number}
\colorbox{blue!10.1882278879446}{of}
\colorbox{blue!10.3796033994334}{different}
\colorbox{blue!16}{price}
\colorbox{blue!7.0758577274158}{brackets}
\colorbox{blue!11.1451054453887}{.}
\colorbox{blue!12.4343720491029}{for}
\colorbox{blue!9.25149512118351}{the}
\colorbox{blue!9.1910607491344}{brand}
\colorbox{blue!5.34340572867485}{called}
\colorbox{blue!12.1926345609065}{ae}
\colorbox{blue!12.3336480956878}{\#\#nd}
\colorbox{blue!12.6559647466163}{\#\#less}
\colorbox{blue!15.1740635819956}{pools}
\colorbox{blue!6.25999370475291}{\#\#a}
\colorbox{blue!14.6301542335537}{prices}
\colorbox{blue!8.36512433113}{start}
\colorbox{blue!10.0371419578218}{at}
\colorbox{blue!14.4488511174064}{\$}
\colorbox{blue!9.39250865596475}{23}
\colorbox{blue!12.5552407932011}{,}
\colorbox{blue!6.74346868114574}{900}
\colorbox{blue!12.4343720491029}{for}
\colorbox{blue!8.09316965690903}{their}
\colorbox{blue!6.81397544853635}{most}
\colorbox{blue!5.84702549575071}{basic}
\colorbox{blue!14.0056657223796}{pool}
\colorbox{blue!12.4343720491029}{for}
\colorbox{blue!9.05004721435316}{other}
\colorbox{blue!9.65439093484419}{brands}
\colorbox{blue!10.1882278879446}{of}
\colorbox{blue!14.7510229776519}{endless}
\colorbox{blue!14.0056657223796}{pool}
\colorbox{blue!12.2429965376141}{(}
\colorbox{blue!10.8530059804847}{a}
\colorbox{blue!11.1451054453887}{.}
\colorbox{blue!0}{k}
\colorbox{blue!11.1451054453887}{.}
\colorbox{blue!10.8530059804847}{a}
\colorbox{blue!8.8385269121813}{swim}
\colorbox{blue!12.3537928863708}{spa}
\colorbox{blue!12.5451683978596}{\#\#s}
\colorbox{blue!2.81523449795404}{)}
\colorbox{blue!14.6301542335537}{prices}
\colorbox{blue!12.2228517469311}{can}
\colorbox{blue!5.6052880075543}{be}
\colorbox{blue!10.0472143531634}{as}
\colorbox{blue!9.16084356310985}{low}
\colorbox{blue!10.0472143531634}{as}
\colorbox{blue!14.4488511174064}{\$}
\colorbox{blue!6.74346868114574}{900}
\colorbox{blue!6.35064526282657}{\#\#0}
\colorbox{blue!9.88605602769909}{things}
\colorbox{blue!10.4903997481901}{that}
\colorbox{blue!7.88164935473717}{change}
\colorbox{blue!9.25149512118351}{the}
\colorbox{blue!16}{price}
\colorbox{blue!14.6301542335537}{prices}
\colorbox{blue!7.88164935473717}{change}
\colorbox{blue!2.03966005665722}{drastically}
\colorbox{blue!4.07428391564369}{depending}
\colorbox{blue!1.52596789423985}{upon}
\colorbox{blue!2.95624803273528}{location}
\colorbox{blue!12.5552407932011}{,}
\colorbox{blue!4.42681775259679}{time}
\colorbox{blue!10.1882278879446}{of}
\colorbox{blue!6.29021089077746}{year}
\colorbox{blue!12.5552407932011}{,}
\colorbox{blue!10.1076487252125}{and}
\colorbox{blue!5.3736229146994}{property}
\colorbox{blue!4.15486307837583}{type}
\colorbox{blue!11.1451054453887}{.}

Top expansion terms:
\colorbox{blue!13.9454247403211}{pay}
\colorbox{blue!13.3209039974819}{paid}
\colorbox{blue!12.0615483789739}{cost}
\colorbox{blue!11.8403102297765}{paying}
\colorbox{blue!11.7995694050992}{much}
\colorbox{blue!11.4571401951527}{what}
\colorbox{blue!10.5510436260623}{fee}
\colorbox{blue!10.490190242367}{costs}
\colorbox{blue!9.96674800125905}{thing}
\colorbox{blue!9.7955333962858}{spending}

\end{mdframed}
\smrmf{
\begin{mdframed}

\smrm{docTTTTquery:
\colorbox{blue!16}{pool}
\colorbox{blue!14.3589743589744}{endless}
\colorbox{blue!7.38461538461539}{how}
\colorbox{blue!7.38461538461539}{cost}
\colorbox{blue!6.56410256410256}{much}
\colorbox{blue!4.1025641025641}{price}
\colorbox{blue!3.28205128205128}{doe}
\colorbox{blue!0.41025641025641}{swim}
\colorbox{blue!0}{build}
\colorbox{blue!0}{spa} ...
}

\end{mdframed}
}
\end{flushleft}
\vspace{-1em}
\caption{Relative representation values of terms that appear in a sample document. Alongside terms that appeared in the passage, the top `expansion' terms are also shown.\smrm{ For reference, the most frequent terms produces by docTTTTTquery are also given, weighted by term frequency.}}
\vspace{-0.5em}
\label{fig:dvec_interpret}
\end{figure}

\textbf{Interpretability of representations}. A benefit of our approach is that the dimensions of the representation correspond to terms in the lexicon, allowing the representations to be easily inspected. In Fig.~\ref{fig:qvec_interpret}, we present the relative scores for sample queries from MS-MARCO. We observe that the model is generally able to pick up on the terms that match intuitions of term importance. For instance, (a) gives highest scores to \textit{california}, \textit{aaa} (American Automobile Association), and \textit{tow}. These three terms are good candidates for a keyword-based query with the same query intent. This approach does not necessarily just remove stop words; in (b) \textit{what} is assigned a relatively high score.
We provide an example of document vector importance scores in Fig.~\ref{fig:dvec_interpret}. Because the document vector is dense, the figure only shows the terms that appear directly in the document and other top-scoring terms. Notice that terms related to \textit{price}, \textit{endless}, \textit{pool(s)}, and \textit{cost} are assigned the highest scores. In this case, the expansion of the term \textit{cost} was critical for properly scoring this document, as a relevant query is \textit{cost of endless pools/spas}. \smrm{Although the terms that docTTTTTquery generate for this document are similar, the continuous values generated by EPIC paid off in a higher MRR@10 score for the query ``\textit{cost of endless pools/swim spa}'' (a relevant question for this passage).}

\section{Conclusion}

We demonstrated an effective and inexpensive technique for re-ranking passages based on lexicon-grounded representations generated from contextualized language models. This work advances the art by further approaching fully BERT-based re-ranking performance, while providing low query-time latency and easy interpretability of representations. We also find that pruning can be an effective technique for reducing query latency \smrm{and the size of the pre-computed passage representations} without sacrificing effectiveness.\smrm{Future work can investigate how well this approach generalizes to document retrieval.}

\section*{Acknowledgments}
Work partially supported by the ARCS Foundation. Work partially supported by the Italian Ministry of Education and Research (MIUR) in the framework of the CrossLab project (Departments of Excellence). Work partially supported by the BIGDATAGRAPES project funded by the EU Horizon 2020 research and innovation programme under grant agreement No. 780751, and by the OK-INSAID project funded by the Italian Ministry of Education and Research (MIUR) under grant agreement No. ARS01\_00917.

\bibliographystyle{ACM-Reference-Format}
\bibliography{biblio}

\end{document}